\pdfoutput=1

\documentclass[reqno]{amsart}
\RequirePackage[l2tabu, orthodox]{nag} % turn on warnings because of bad style

\usepackage{hyperref}

\usepackage{courier}
\usepackage{color}
\usepackage{amsmath,amsthm,amssymb}
\usepackage{cases} %numcases
\usepackage[cal=boondox,scr=boondoxo]{mathalfa}

\usepackage{graphicx}
\usepackage[figurename=Fig.,labelsep=period]{caption}

\usepackage{textcomp} % TS1 encoding (upright Greek mu)
\usepackage{sistyle}
\usepackage{physics}
\usepackage{mathtools}
\mathtoolsset{showonlyrefs=false}
\theoremstyle{definition}
\newtheorem{remark}{Remark}

\usepackage[]{natbib}
\bibpunct[,]{(}{)}{,}{a}{}{,}

\begin{document}

\title[Hatchery-induced transition in a Pareto population]{Hatchery-induced transition of the effective size in a Pareto population}
\author[H.-S. Niwa]{H.-S. Niwa}
%\date{\today}
\date{}


\begin{abstract}
It seems paradoxical to have observed the absence of reduced effective population sizes $N_{\mathrm{e}}$ under marine hatchery practices. This paper studies the Ryman-Laikre, or two-demographic-component, model of the hatchery impact related to inbreeding in a population with power-law family-size distribution, where hatchery inputs are represented by a Dirac delta function. By examining the asymptotic (i.e. large-population limit) behavior of the normalized sizes (or weights) of families of the mixture population, I derive the distribution properties of the average weight of families (i.e. the sum of the squared weights, $Y$) over the population existing at any given time. The reciprocal of the average weight $Y$ gives the effective number of families (or reproducing lineages) in the population, $N_{\mathrm{e}}=1/Y$. When the specific production in the hatchery (i.e. the number of offspring per broodstock) is low, the most probable value of $N_{\mathrm{e}}$ is close to the lower bound of the $N_{\mathrm{e}}$-distribution. When the specific hatchery-production is increased to a critical value with fixed mixing proportion of hatchery fish, a discontinuous transition takes place, so that the most probable $N_{\mathrm{e}}$ jumps to the upper extreme of the distribution. This hatchery-induced transition is attributed to the breaking of reciprocal symmetry, i.e. the fact that the typical value of $Y$ and its reciprocal (the typical $N_{\mathrm{e}}$) do not vary with the population size in opposite ways. At a high specific hatchery-production, the symmetry breaking disappears.
\end{abstract}

\maketitle

\section{Introduction}
Stocking or supplementation of aquatic systems with hatchery-reared juveniles is a common management practice in fisheries.
Large scale industrial releases of hatchery-produced fish have been conducted
since the mid-nineteenth century
\citep{Lorenzen2005}.
Worldwide, 64 countries reported some hatchery activity in marine and coastal stocking with approximately 180 different species being released by the late 1990s
\citep{Born-etal2004}.
The hatchery impact related to inbreeding is particularly important for highly fecund marine species with type-III survivorship curve.
Hatchery practices for mass spawning species collect a small number of broodstock from the wild.
Hatchery-rearing enhances survival of early life stages and millions of hatchery-reared fish are released into the wild.
While the hatchery breeders will represent a tiny fraction of the population, e.g. 0.01\% or lower
\citep{Waples-etal2016},
their offspring (hatchery fish) can make up a substantial fraction of the next generation.

\citet{Ryman-Laikre91} addressed the effect of hatchery stocking on the effective size of an admixed population,
which is composed of two demographic components,
one in captivity and the other in the native setting.
They provided the formula
\begin{equation}\label{eqn:RL-formula}
 \frac{1}{N_{\mathrm{e}}}=
  \frac{p^2}{N_{\mathrm{e}}^{\scalebox{0.55}{(1)}}} + \frac{(1-p)^2}{N_{\mathrm{e}}^{\scalebox{0.55}{(2)}}}
\end{equation}
for total effective number ($N_{\mathrm{e}}$) of the hatchery and wild parents in combination, where $p$ is the mixing weight of offspring from the hatchery ($0<p<1$),
and $N_{\mathrm{e}}^{(i)}$ refers to the effective number of parents of hatchery ($i=1$) or wild ($i=2$) component of the population.
It is anticipated that hatchery stocking causes reduced $N_{\mathrm{e}}$ compared to unsupplemented demography
(i.e. Ryman-Laikre effect).
Paradoxically, some studies have reported the absence of reduced $N_{\mathrm{e}}$ for marine populations under hatchery stocking
\citep[e.g.][]{Carson-etal2009,Nakajima-Kitada-cjfas2014}.
There was, however, observed a high variability in $N_{\mathrm{e}}$ among years in the seeded great scallop \textit{Pecten maximus} population of the Bay of Brest, France
\citep{Morvezen-heredity2016}.

Very large interfamilial, or sweepstakes, variation in reproductive success has been documented in abundant marine species with type-III survivorship curve
\citep{Hedgecock-Pudovkin2011}.
A Pareto (power-law) offspring-number distribution naturally arises from type-III (exponential) survivorship with family-correlated survival
\citep{Reed-Hughes2002,Niwa-etal2017}.
By contrast, the commercial hatchery consistently and systematically produces millions of juveniles;
it is crucial to have high hatchery productivity and low variation in offspring number among broodstock fish.

This paper is motivated by the puzzling absence of reduced $N_{\mathrm{e}}$ under marine hatchery practices,
and studies the distribution properties of normalized sizes (or weights) of families in a Pareto population with hatchery inputs represented by a Dirac delta function.
I address the asymptotic behavior of the model in the large-population limit;
it serves as an approximation for large population size.
Letting $Y$ be the sum of the squared weights of families,
its reciprocal gives the effective number of families (or reproducing lineages) in the population,
$N_{\mathrm{e}}{\;}(=Y^{-1})$.
Examining the probability distribution of $N_{\mathrm{e}}$,
I show that an increased individual-specific (or per-capita) production of hatchery fish triggers a drastic change in the most probable $N_{\mathrm{e}}$
from the lower to the upper extreme of the distribution.

\section{Pareto population with hatchery inputs}
In this section, after presenting a mathematical model of the aquatic systems with hatchery inputs,
I give an overview of some general results on the statistical domination by the largest term in the sum of independent Pareto random variables.
Then, I compute the size-frequency distribution of families, $\rho(w)$,
which is defined in the following way:
$\rho(w)\dd{w}$ is the mean number of families with weights between $w$ and $w+\dd{w}$ among an infinite number of replicate populations each undergoing the same reproduction process.
From the distribution $\rho(w)$ together with the concept of the domination by the largest term,
the asymptotic expressions for the probability distributions of $Y$ and $N_{\mathrm{e}}$ are derived in \S\ref{section:distribution-Y2}, which is the core of the paper.

\subsection{Sums of random variables}
The marine population with hatchery inputs is modeled as a two-component mixture.
The hatchery and wild components of the population are designated by labels~1 and~2, respectively.
Consider a population consisting of $N$ haploid individuals,
where fixed numbers $N_1$ and $N_2$ of them have reproduction condition~1 and~2, respectively ($N_1+N_2=N$).
In a given generation, each individual has an equal chance of being assigned to each condition,
and individual $j\,(=1,\ldots,N_i)$ in condition $i\,(=1,2)$ is assigned
a random value $X_j^{(i)}$ of reproductive success (a positive real number),
drawn independently from a probability density $f_i(x)$.
The component-1 density is a Dirac delta function, $f_1(x)=\delta(x-u)$, concentrated at $u>0$.
The component-2 density is a $\mathrm{Pareto}(\alpha)$ distribution with $1<\alpha<2$,
\begin{equation}\label{eqn:density-2}
 f_2(x) = \alpha x^{-\alpha-1}
\end{equation}
on $x\geq 1$,
where the mean $\mu=\alpha/(\alpha-1)$ is finite
and the second moment is infinite.
Upon normalizing $X_j^{(i)}$'s ($i=1,2;\ j=1,\ldots,N_i$) by their sum
\begin{equation*}
 R_N=uN_1+\sum_{j=1}^{N_2} X_j^{(2)},
\end{equation*}
one defines the weight $W_j^{(i)}$ of the term $X_j^{(i)}$ in the sum as
\begin{equation*}
 W_j^{(i)}=X_j^{(i)}/R_N.
\end{equation*}
Each weight $W_j^{(i)}$ gives the probability of reproductive success of individual $j$ in condition $i$,
so the $j$-th family in component $i$ recruits a fraction $W_j^{(i)}$ of the next generation.
To put it another way,
given the population at some generation, for each individual at the following generation, one chooses at random with probability $W_j^{(i)}$ one parent $j\in\{1,\ldots,N_i\}$ in component $i\in\{1,2\}$.
Any generation is replaced by a new one.
The mixing weight of hatchery fish, $uN_1/R_N$, is in the mean (or in the limit $N\to\infty$)
\begin{equation*}
 p = \qty(1+\frac{\mu N_2}{uN_1})^{-1},
\end{equation*}
which is assumed to be constant and independent of the population size, i.e.
$p=\order{1}$ in $1/N$.
The asymptotic behavior of two-component mixtures is examined
under the assumption
\begin{equation*}%\label{eqn:uN1N2-ratio}
 \left\{
  \begin{aligned}
   uN_1 &= \order{N}\\
   N_2 &= \order{N}
  \end{aligned}
 \right.
\end{equation*}
with $\mu=\order{1}$.

When choosing an individual at random,
the probability that it is in the $j$-th family in component $i$ is $W_j^{(i)}$,
so the sum
\begin{equation*}
 Y=\sum_{i=1}^2\sum_{j=1}^{N_i}\qty(W_j^{(i)})^2
\end{equation*}
gives the expected weight of the family containing it.
The reciprocal of this average weight $Y$ for a realization of the reproduction process gives the effective number of families in the population, i.e.
the effective population size $N_{\mathrm{e}}=Y^{-1}$ at a given generation.

The $X_j^{(i)}$ gives an analog of the number of potential offspring (i.e. surviving young to reproductive maturity) of individual $j$ in condition~$i$, and
$u$ is the individual-specific production of young fish (i.e. the number of offspring per broodstock) in the hatchery.
The sum $R_N$ approximates the total number of recruits entering the (potentially reproductive) population.
$N$ is the number of reproducing individuals in the population.
Note that
the above is just the nest-site model \citep{Wakeley09} but with reproduction laws being qualitatively different at the two nest-sites.

\subsection{Domination by the largest term}
\noindent
For the results presented in this subsection, I follow
\citet{Bouchaud-Georges90}, \citet{ZKS05} and \citet{Hofstad2016}.
Write $R_{N_2}^{(2)}$ for the sum of $N_2$ independent random variables $X_1^{(2)},\ldots,X_{N_2}^{(2)}$ drawn from the $\mathrm{Pareto}(\alpha)$ distribution (Equation~\ref{eqn:density-2}) with $1<\alpha<2$, i.e.
$R_{N_2}^{(2)}=\sum_{j=1}^{N_2} X_j^{(2)}$.
It is well known that
$(R_{N_2}^{(2)}-\mu N_2)/N_2^{1/\alpha}$ has an $\alpha$-stable distribution for large $N_2$.
The width of the distribution of the sum $R_{N_2}^{(2)}$ (i.e. the typical value of the difference $R_{N_2}^{(2)}-\mathrm{E}[R_{N_2}^{(2)}]$) is of $\order{N_2^{1/\alpha}}$,
while the variance $\mathrm{E}[(R_{N_2}^{(2)}-\mathrm{E}[R_{N_2}^{(2)}])^2]$ is infinite.
$\mathrm{E}[\,\cdot\,]$ represents the mean over all possible draws.

Define $X_{1,N_2}^{(2)}\geq X_{2,N_2}^{(2)}\geq\cdots\geq X_{N_2,N_2}^{(2)}$ by ranking in decreasing order the values encountered among the $N_2$ terms of the sum $R_{N_2}^{(2)}$.
The typical value of the largest observation $X_{1,N_2}^{(2)}$ grows as $N_2^{1/\alpha}$,
or more precisely,
the rescaled random variable
$X_{1,N_2}^{(2)}/N_2^{1/\alpha}$
also has a $\mathrm{Pareto}(\alpha)$ distribution
with
\begin{equation*}
 \mathrm{E}\qty[X_{1,N_2}^{(2)}/N_2^{1/\alpha}]
  =\mathrm{\Gamma}(1-1/\alpha)
\end{equation*}
for large $N_2$.
While $X_{1,N_2}^{(2)}$ has an infinite second moment, one has
\begin{align*}
 \mathrm{E}\qty[X_{2,N_2}^{(2)}]&=\frac{\alpha-1}{\alpha}\mathrm{E}\qty[X_{1,N_2}^{(2)}]\\
 \mathrm{E}\qty[\qty(X_{2,N_2}^{(2)})^2]&=\frac{N_2!{\,}\mathrm{\Gamma}(2-2/\alpha)}{\mathrm{\Gamma}(N_2+1-2/\alpha)}
 =\mathrm{\Gamma}(2-2/\alpha)N_2^{2/\alpha}
 \end{align*}
for large $N_2$.
One also has
\begin{equation*}
 \mathrm{E}\qty[\sum_{j=2}^{N_2}\qty(X_{j,N_2}^{(2)})^2] =
  \frac{\alpha}{\alpha-2}
  \qty(N_2-\mathrm{\Gamma}(2-2/\alpha)N_2^{2/\alpha})
  =\frac{\alpha}{2-\alpha}\mathrm{E}\qty[\qty(X_{2,N_2}^{(2)})^2].
\end{equation*}
Importantly, all but the largest order statistics have finite second moment.
Accordingly, the statistical variation of the sum $R_{N_2}^{(2)}$ is dominated by its largest term $X_{1,N_2}^{(2)}$.
So, unless $u^2N_1\gtrsim N^{2/\alpha}$
(i.e. the specific hatchery-production is sufficiently high, $u\gtrsim N^{2/\alpha-1}$),
the fraction of the sum, $(R_N-\mathrm{E}[R_N])/R_N$, can be linked to one parent in component~2.

\subsection{Size-frequency distribution of families}
Consider the distribution of the weights of families
\begin{equation*}
 \rho(w) = \mathrm{E}\qty[\sum_{i=1}^2\sum_{j=1}^{N_i}\delta\qty(w-W_j^{(i)})],
\end{equation*}
which is the mean of the empirical distribution of the number of families.
The mean is taken over all possible partitions of the unit interval to $N$ parts with measures $\{W_j^{(i)}\mid i=1,2;\ j=1,\ldots,N_i\}$.
The probability of a randomly-chosen individual coming from a family of weight $w$ (i.e. the probability of finding a family of weight $w$) is given by $w\rho(w)$.
Write $\tilde{W}_j^{(2)}$ for the weight of the term $X_j^{(2)}$ in the sum $R_{N_2}^{(2)}$ as
$\tilde{W}_j^{(2)}=X_j^{(2)}/R_{N_2}^{(2)}$.
Letting $\rho_2(w)$ be the number of families in component~2 at weight $w$,
\begin{equation*}
 \rho_2(w) = \mathrm{E}\qty[\sum_{j=1}^{N_2}\delta\qty(w-\tilde{W}_j^{(2)})],
\end{equation*}
the function $\rho_2(w)$ is extracted as
\begin{equation}\label{eqn:freq-spectrum-beta}
 c_2^{-1}w^2\rho_2(w)=
  \frac{\alpha w^{1-\alpha}(1-w)^{\alpha-1}}{\mathrm{Beta}(2-\alpha,\alpha)}
\end{equation}
with
\begin{equation}\label{eqn:multi-merger}
 c_2=
  \frac{\alpha\mathrm{Beta}(2-\alpha,\alpha)}{\mu^{\alpha} N_2^{\alpha-1}}
\end{equation}
in the limit $N_2\to\infty$
\citep{Niwa-arxiv10Feb2022},
where
$\mathrm{Beta}(a,b)=\mathrm{\Gamma}(a)\mathrm{\Gamma}(b)/\mathrm{\Gamma}(a+b)$ is the beta function.
Letting $q{\;}(=1-p)$ be the mixing weight of component~2,
\begin{equation*}
 q=\lim_{N\to\infty}R_{N_2}^{(2)}/R_N,
\end{equation*}
the asymptotic distribution of the weights of families reduces to
\begin{align}\label{eqn:rho-mixture}
 \rho(w) &=
 N_1\delta\qty(w-\frac{p}{N_1})+
 \mathrm{E}\qty[\sum_{j=1}^{N_2}\delta\qty(w-q\tilde{W}_j^{(2)})]\nonumber\\
 &=
 N_1\delta\qty(w-\frac{p}{N_1})+
 q^{-1}\rho_2\qty(\frac{w}{q})\Theta\qty(1-\frac{w}{q})
\end{align}
with Heaviside step function $\Theta(\cdot)$.
Write $\varepsilon_{N_2}$ for a minimum weight of families (i.e. a cut-off in the region of small $w$) in component~2,
\begin{equation*}
 \varepsilon_{N_2}=(\mu N_2)^{-1}.
\end{equation*}
Then, one has
\begin{equation*}
 \lim_{N_2\to\infty}
  N_2^{-1}\int_{\varepsilon_{N_2}}^1 \rho_2(w)\dd{w}=1
\end{equation*}
and
\begin{equation*}
 \lim_{N_2\to\infty}
  \int_{\varepsilon_{N_2}}^1 w\rho_2(w)\dd{w}=1,
\end{equation*}
which ensure both that $\rho(w)$ integrates to $N$ and that $w\rho(w)$ integrates to unity on $q\varepsilon_{N_2}\leq w\leq 1$ for large $N$
(where $q\varepsilon_{N_2}<p/N_1$ is assumed).

The asymptotic behavior of $\mathrm{E}[Y]$ in two-component mixtures will depend on the relative order of magnitude of $N_1$ and $N_2$.
If
\begin{equation}\label{eqn:N1N2-ratio}
 N_1 = \order{N^{\alpha-1}},
\end{equation}
Equation~\eqref{eqn:rho-mixture} yields for large $N$
\begin{equation}\label{eqn:nest-site-coal-probab}
 c_{\scalebox{0.55}{RL}}\equiv \mathrm{E}\qty[Y] =\int_0^1 w^2\rho(w)\dd{w}
  = p^2/N_1 + q^2 c_2
\end{equation}
with $c_2$ as in Equation~\eqref{eqn:multi-merger},
which implies that
the Ryman-Laikre formula (Equation~\ref{eqn:RL-formula}) holds for the mean values over all realizations of the reproduction process.

%{Remark 1}
\begin{remark}
If one assumes the component-1 density to be a gamma distribution
\begin{equation}\label{eqn:density-1}
 f_1(x)=\frac{x^{u-1}e^{-x}}{\mathrm{\Gamma}(u)}
\end{equation}
($x>0$)
with variance equal to the mean $u>0$,
and considers the weight of the term $X_j^{(1)}$ in the sum $\sum_{j=1}^{N_1} X_j^{(1)}$,
where $X_j^{(1)}$'s are independently distributed according to Equation~\eqref{eqn:density-1},
then the number of families (in component~1) at weight $w$ has the distribution
\begin{equation*}
 \rho_1(w)=\frac{N_1 w^{u-1}(1-w)^{(N_1-1)u-1}}{\mathrm{Beta}\qty(u,(N_1-1)u)}
\end{equation*}
\citep{Cramer45},
which for large $N_1$ (or large $u$) approximates the Dirac delta function,
$\rho_1(w) = N_1\delta(w-N_1^{-1})$.
\end{remark}

%{Remark 2}
\begin{remark}
Since all the moments of weights, conditional on the component-2 density, are known as
\begin{equation*}
 c_2^{-1}\mathrm{E}\qty[\sum_{j=1}^{N_2}\qty(\tilde{W}_j^{(2)})^k]=
  \frac{\mathrm{Beta}(k-\alpha,\alpha)}{\mathrm{Beta}(2-\alpha,\alpha)}
\end{equation*}
($k\geq 2$)
in the limit $N_2\to\infty$
(\citet{Huillet2014,Niwa-arxiv10Feb2022}; see also online supplementary appendix D of \citealt{Niwa-etal2016}),
the distribution $c_2^{-1}w^2\rho_2(w)$ is known
\citep{Hausdorff1923},
which is the $\mathrm{Beta}(2-\alpha,\alpha)$ density on $[0,1]$ as in Equation~\eqref{eqn:freq-spectrum-beta}.
\end{remark}

%{Remark 3}
\begin{remark}
When the offspring-number distribution has a power-law tail with $1<\alpha<2$, asynchronous, arbitrary multiple collisions (i.e. $\mathrm{Beta}(2-\alpha,\alpha){\ }\Lambda$-coalescents) of ancestral lineages occur
\citep{Schweinsberg2003}.
For the two-component population model with $\lim_{N\to\infty} N_1^{-1}=0$, the distribution $w^2\rho(w)$ has a Dirac mass at $w=0$ in the large-$N$ limit.
Under the scaling of Equation~\eqref{eqn:N1N2-ratio}, the presence of a Dirac mass at zero adds a Kingman component to the $\mathrm{\Lambda}$-coalescent
\citep{Berestycki2009}.
In addition to multiple collisions governed by $q^{-1}\rho_2(w/q)$, each pair of lineages coalesces at rate $p^2 N_1^{-1}c_{\scalebox{0.55}{RL}}^{-1}$, with $c_{\scalebox{0.55}{RL}}$ as in Equation~\eqref{eqn:nest-site-coal-probab}.
\end{remark}

\section{Distribution properties of the average weight}\label{section:distribution-Y2}
In this Section,
the expressions for the probability distributions of $Y$ and $N_{\mathrm{e}}$ are obtained in the case when the hatchery productivity is low such that
\begin{equation}\label{eqn:low-mixing}
 u^2N_1\lesssim\mathrm{E}\qty[X_{1,N_2}^{(2)}]^2
  \qquad
  \qty(\mbox{i.e.{\ \ }} u\lesssim N^{2/\alpha-1})
\end{equation}
and in the high-productivity case
\begin{equation*}
 u\gtrsim N^{2/\alpha-1}.
\end{equation*}
These probability distributions are also obtained numerically.

\subsection{Low specific hatchery-production}
Following \citet{Derrida-Flyvbjerg87-PhysA},
I first compute the probability distribution $\mathrm{\Pi}_{W_{1,N}}(w)$ of the weight $W_{1,N}$ of the largest family.
If there is a family with weight $w>1/2$, this weight must be the largest one, and thus
$\mathrm{\Pi}_{W_{1,N}}(w)=\rho(w)$ for $w>1/2$.
Letting $\rho^{\ast}(w_1,\ldots,w_n)$ be a joint distribution of weights of $n$ different families ($n\geq 2$),
and denoting
\begin{equation*}
 I_n(w)=\int_w^1\dd{v_1}\int_w^{v_1}\dd{v_2}\cdots\int_w^{v_{n-2}}\dd{v_{n-1}}\rho^{\ast}(v_1,v_2,\ldots,v_{n-1},w),
\end{equation*}
one has,
in the interval $1/(n+1)<w<1/n$,
\begin{equation*}
 \mathrm{\Pi}_{W_{1,N}}(w)=\rho(w)-I_2(w)+\cdots+(-1)^{n-1}I_n(w).
\end{equation*}
The successive terms in these summations rapidly decrease for large $N$,
because $I_n(w)$ is an integral over more and more variables and diminishes in magnitude with increasing $n$.
In the low-productivity case of Equation~\eqref{eqn:low-mixing},
the weights of families from the hatchery,
\begin{equation*}
 u/R_N\approx p/N_1{\;}
  \qty(\lesssim N^{2(1/\alpha-1)}),
\end{equation*}
are less than the typical weight of the largest family in component~2,
\begin{equation*}
 \hat{w}_N=N_2^{1/\alpha-1}\mu^{-1}.
\end{equation*}
Accordingly, in the large-$N$ limit, the largest family is seen in condition~2, and one has, up to the leading term,
\begin{equation}\label{eqn:largest-W-large-N}
 \mathrm{\Pi}_{W_{1,N}}(w) = q^{-1}\rho_2(w/q)
\end{equation}
on $N^{1/\alpha-1}\lesssim w\leq q$.
Note that integrating Equation~\eqref{eqn:largest-W-large-N} gives
\begin{equation*}
 \int_{q\hat{w}_N}^{q} \mathrm{\Pi}_{W_{1,N}}(w)\dd{w}=1.
\end{equation*}
One also sees that
\begin{equation*}
 \mathrm{E}[W_{1,N}]=
  \int_{q\hat{w}_N}^{q} w\mathrm{\Pi}_{W_{1,N}}(w)\dd{w}
  = qN_2^{1/\alpha-1}.
\end{equation*}

The probability distribution $\mathrm{\Pi}_{Y}(y)$ of $Y$ is defined as
\begin{equation*}
 \mathrm{\Pi}_{Y}(y)=\mathrm{E}\qty[\delta\qty(
 \sum_{i=1}^{2}\sum_{j=1}^{N_i}\qty(W_j^{(i)})^2 -y)].
\end{equation*}
Since the sums which contribute to the mean are those in which one term dominates,
following \citet{MezardPSTV84},
one may replace the sum with its maximal summand,
\begin{equation*}
 \mathrm{\Pi}_{Y}(y) = \mathrm{E}\qty[\delta\qty(W_{1,N}^2-y)].
\end{equation*}
From Equation~\eqref{eqn:largest-W-large-N}, one obtains
\begin{align}\label{eqn:distr-Y2}
 \mathrm{\Pi}_{Y}(y) &=
  \int_{q\hat{w}_N}^{q}
  \mathrm{\Pi}_{W_{1,N}}(w)\delta\qty(w^2-y)\dd{w}\nonumber\\
 &=
 \frac{\rho_2\qty(\sqrt{y/q^2})}{2q\sqrt{y}}
 =
  \frac{\alpha \qty(y/q^2)^{-\alpha/2-1}\qty(1-\sqrt{y/q^2})^{\alpha-1}}
  {2q^2 \mu^{\alpha} N_2^{\alpha-1}}
\end{align}
on $N^{2(1/\alpha-1)}\lesssim y\leq q^2$.
Note that integrating Equation~\eqref{eqn:distr-Y2} gives
\begin{equation*}
 \int_{(q\hat{w}_N)^2}^{q^2} \mathrm{\Pi}_{Y}(y)\dd{y}=1.
\end{equation*}
One sees that
the most probable value of the $Y$ is close to the typical value of squared weight of the largest family in component~2, and is very different from the mean value
\begin{equation}\label{eqn:mean-Y2}
 \mathrm{E}[Y] = \int_{(q\hat{w}_N)^2}^{q^2} y{\,}\mathrm{\Pi}_{Y}(y)\dd{y}
  = q^2 c_2
\end{equation}
with $c_2$ as in Equation~\eqref{eqn:multi-merger}.

The same kind of argument applies to the probability distribution of $N_{\mathrm{e}}{\;}(=Y^{-1})$,
yielding
\begin{align}\label{eqn:distr-Ne}
 \mathrm{\Pi}_{N_{\mathrm{e}}}(y) &=
  \int_{q\hat{w}_N}^{q}
  \mathrm{\Pi}_{W_{1,N}}(w)\delta\qty(w^{-2}-y)\dd{w}\nonumber\\
 &=
 \frac{\rho_2\qty(1/\sqrt{q^2y})}{2qy^{3/2}}
 =
 \frac{
 q^2\alpha\qty(q^2 y)^{\alpha/2-1}
  \qty(1-1/\sqrt{q^2 y})^{\alpha-1}
 }{2\mu^{\alpha} N_2^{\alpha-1}}
\end{align}
on $q^{-2}\leq y\lesssim N^{2(1-1/\alpha)}$.
The distribution $\mathrm{\Pi}_{N_{\mathrm{e}}}(y)$ is cut off at around the reciprocal of the typical $Y$.
Integrating Equation~\eqref{eqn:distr-Ne} gives
\begin{equation*}
 \int_{q^{-2}}^{(q\hat{w}_N)^{-2}}
  \mathrm{\Pi}_{N_{\mathrm{e}}}(y)\dd{y}=1.
\end{equation*}

Notice that the $Y$-distribution is of reciprocal symmetry breaking,
in the sense that
the typical value and its reciprocal do not vary with population size in opposite ways
\citep{Romeo2003,Niwa-arxiv10Feb2022}.
The typical $Y$ decreases with the population size,
while the typical reciprocal of $Y$ is independent of $N$, given by $N_{\mathrm{e}}\simeq q^{-2}$.

\subsection{Critical specific hatchery-production}
When the hatchery production becomes large such that
\begin{equation*}
 \sum_{j=1}^{N_1}\qty(W_j^{(1)})^2\approx\frac{p^2}{N_1}\gtrsim
  \mathrm{E}\qty[{X_{1,N_2}^{(2)}}/{R_N}]^2\approx
  q^2\hat{w}_N^2
  =\order{N^{2(1/\alpha-1)}},
\end{equation*}
the $Y$-distribution is cut off at $y=p^2/N_1+q^2\hat{w}_N^2$.
In the case $p/N_1\lesssim q\hat{w}_N$ (i.e. $u\lesssim \mathrm{E}[X_{1,N_2}^{(2)}]$),
from Equation~\eqref{eqn:largest-W-large-N} one gets
\begin{equation}\label{eqn:distr-Y2-cr}
 \mathrm{\Pi}_{Y}(y) =
 \int_{q\hat{w}_N}^{q}
 \mathrm{\Pi}_{W_{1,N}}(w)\delta\qty(\frac{p^2}{N_1}+w^2-y)\dd{w}
 = \frac{\rho_2\qty(q^{-1}\sqrt{y-p^2/N_1})}{2q\sqrt{y-p^2/N_1}}
\end{equation}
on $p^2/N_1+q^2\hat{w}_N^2\leq y\leq q^2$.
The mean of $Y$ is given by
\begin{equation*}
 \mathrm{E}[Y] = \int_{p^2/N_1+q^2\hat{w}_N^2}^{q^2} y{\,}\mathrm{\Pi}_{Y}(y)\dd{y}
 =
 \frac{p^2}{N_1} + q^2 c_2,
\end{equation*}
which coincides with Equation~\eqref{eqn:nest-site-coal-probab}
or the Ryman-Laikre formula.

One also gets
\begin{equation}\label{eqn:distr-Ne-cr}
 \mathrm{\Pi}_{N_{\mathrm{e}}}(y) =
  \frac{\rho_2\qty(q^{-1}\sqrt{y^{-1}-p^2/N_1})}{2qy^2\sqrt{y^{-1}-p^2/N_1}}
\end{equation}
on $q^{-2}\leq y\leq (p^2/N_1+q^2\hat{w}_N^2)^{-1}$.
The distribution $\mathrm{\Pi}_{N_{\mathrm{e}}}(y)$ is cut off at $y=(p^2/N_1+q^2\hat{w}_N^2)^{-1}$.
One can check the normalization of distributions $\mathrm{\Pi}_{Y}(y)$ and $\mathrm{\Pi}_{N_{\mathrm{e}}}(y)$.

\subsection{Very high specific hatchery-production}
When the specific hatchery-production is very high, such that $u\gtrsim\mathrm{E}[X_{1,N_2}^{(2)}]$,
then the weight of a family
from the hatchery is
$p/N_1\gtrsim q\hat{w}_N$.
The $W_{1,N}$-distribution is cut off at $w=p/N_1$ and
a divergence appears at the lower boundary $p/N_1$ of the $\mathrm{\Pi}_{W_{1,N}}(w)$.
Since Equation~\eqref{eqn:largest-W-large-N} still holds for $w>p/N_1$,
by imposing the normalization condition, one has
\begin{equation}\label{eqn:largest-W-yas}
 \mathrm{\Pi}_{W_{1,N}}(w) =
  \qty(1-\qty(\frac{q\hat{w}_N}{p/N_1})^{\alpha})
  \delta\qty(w-\frac{p}{N_1})
  +q^{-1}\rho_2\qty(\frac{w}{q})
\end{equation}
on $p/N_1\leq w\leq q$,
which yields the probability distribution of $Y$
\begin{align}\label{eqn:distr-Y2-yas}
 \mathrm{\Pi}_{Y}(y) &=
 \int_{p/N_1}^{q}
 \mathrm{\Pi}_{W_{1,N}}(w)\delta\qty(\frac{p^2}{N_1}+w^2-y)\dd{w}\nonumber\\
 &=
 \qty(1-\qty(\frac{q\hat{w}_N}{p/N_1})^{\alpha})
 \delta\qty(y-\frac{p^2}{N_1}-\qty(\frac{p}{N_1})^2)+
 \frac{\rho_2\qty(q^{-1}\sqrt{y-p^2/N_1})}{2q\sqrt{y-p^2/N_1}}
\end{align}
on $p^2/N_1+(p/N_1)^2\leq y\leq q^2$.
The mean of $Y$ is given by
\begin{equation*}
 \mathrm{E}[Y] = \int_{p^2/N_1+(p/N_1)^2}^{q^2} y{\,}\mathrm{\Pi}_{Y}(y)\dd{y}
 =
 \frac{p^2}{N_1} + q^2 c_2
 +\qty(\frac{p}{N_1})^2\qty(1-\qty(\frac{q\hat{w}_N}{p/N_1})^{\alpha})
\end{equation*}
with $N_1$ of order less than $N^{1-1/\alpha}$ and $c_2=\order{N^{1-\alpha}}$.
Therefore, the component~2 makes a negligible contribution to the mean,
and the mean value of $Y$ coincides with its most probable value.

Also from Equation~\eqref{eqn:largest-W-yas} the probability distribution of $N_{\mathrm{e}}$ is extracted as
\begin{align}\label{eqn:distr-Ne-yas}
 \mathrm{\Pi}_{N_{\mathrm{e}}}(y) &=
 \int_{p/N_1}^{q}
 \mathrm{\Pi}_{W_{1,N}}(w)\delta\qty(\qty(\frac{p^2}{N_1}+w^2)^{-1}-y)\dd{w}\nonumber\\
 &=
 \qty(1-\qty(\frac{q\hat{w}_N}{p/N_1})^{\alpha})
 \delta\qty(y-\qty(\frac{p^2}{N_1}+\qty(\frac{p}{N_1})^2)^{-1})+
 \frac{\rho_2\qty(q^{-1}\sqrt{y^{-1}-p^2/N_1})}{2qy^2\sqrt{y^{-1}-p^2/N_1}}
\end{align}
on $q^{-2}\leq y\leq (p^2/N_1+(p/N_1)^2)^{-1}$.
One can check the normalization of distributions $\mathrm{\Pi}_{Y}(y)$ and $\mathrm{\Pi}_{N_{\mathrm{e}}}(y)$.

\subsection{Numerical reconstruction of probability distributions}
I have performed simulations of the two-component population model with hatchery stocking of marine fisheries in mind.
A substantial fraction $p$ of the population is derived from a very small number of hatchery parents ($N_1$ in condition~1).
In the following simulations,
$N_1$ individuals reproduce in condition~1,
where $X_j^{(1)}$'s ($j=1,\ldots,N_1$) are independently drawn from the $\mathrm{gamma}(u)$ distribution in Equation~\eqref{eqn:density-1}
with variance equal to the mean $u=pN_2\mu/(qN_1)$.
$N_2$ individuals reproduce in condition~2, where
$X_j^{(2)}$'s are drawn from the $\mathrm{Pareto}(\alpha)$ distribution in Equation~\eqref{eqn:density-2}.
The distributions of $W_{1,N}$, $Y$ and $N_{\mathrm{e}}$ are generated for a large population ($N=10^5$).

\begin{figure}[hbt!]
\centering
 \begin{tabular}{ll}
  \small{(a)} & \small{(b)}\\
  \includegraphics[height=.3\textwidth,viewport=0 0 360 232]{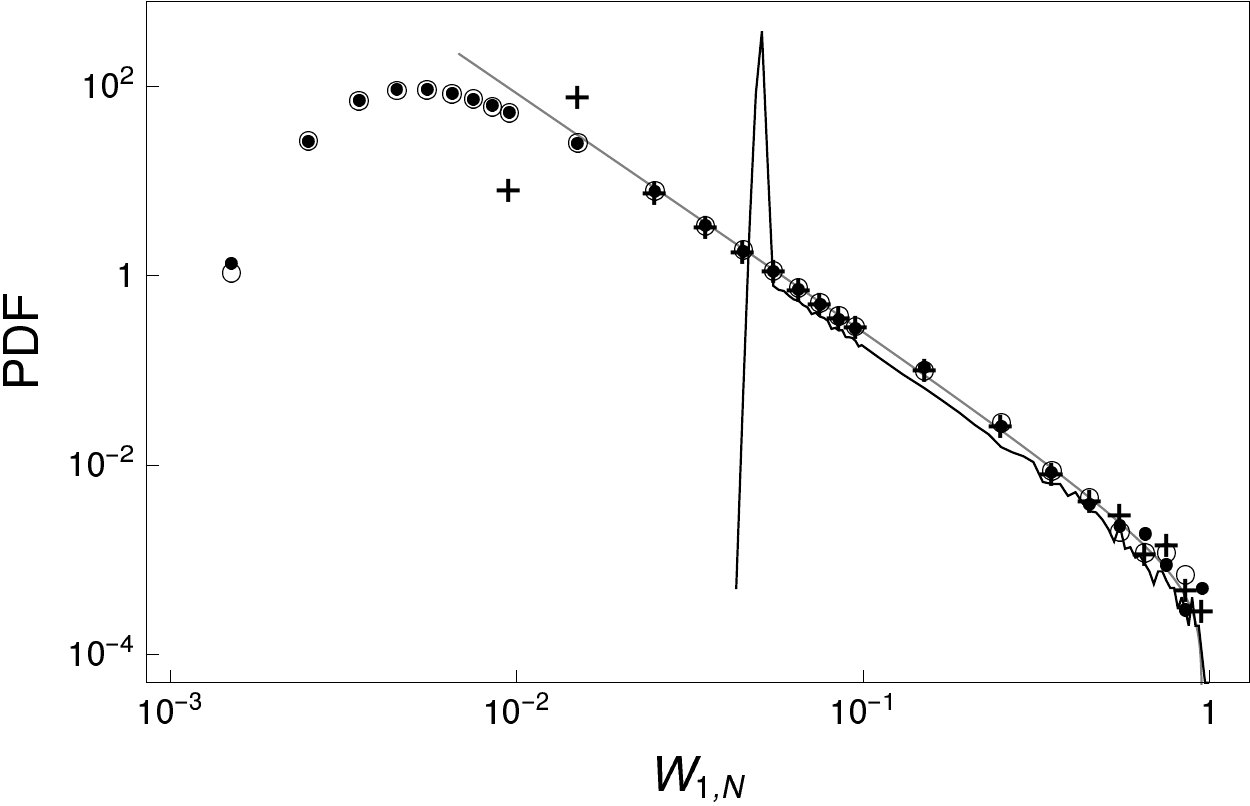}&
  \includegraphics[height=.3\textwidth,viewport=0 0 360 231]{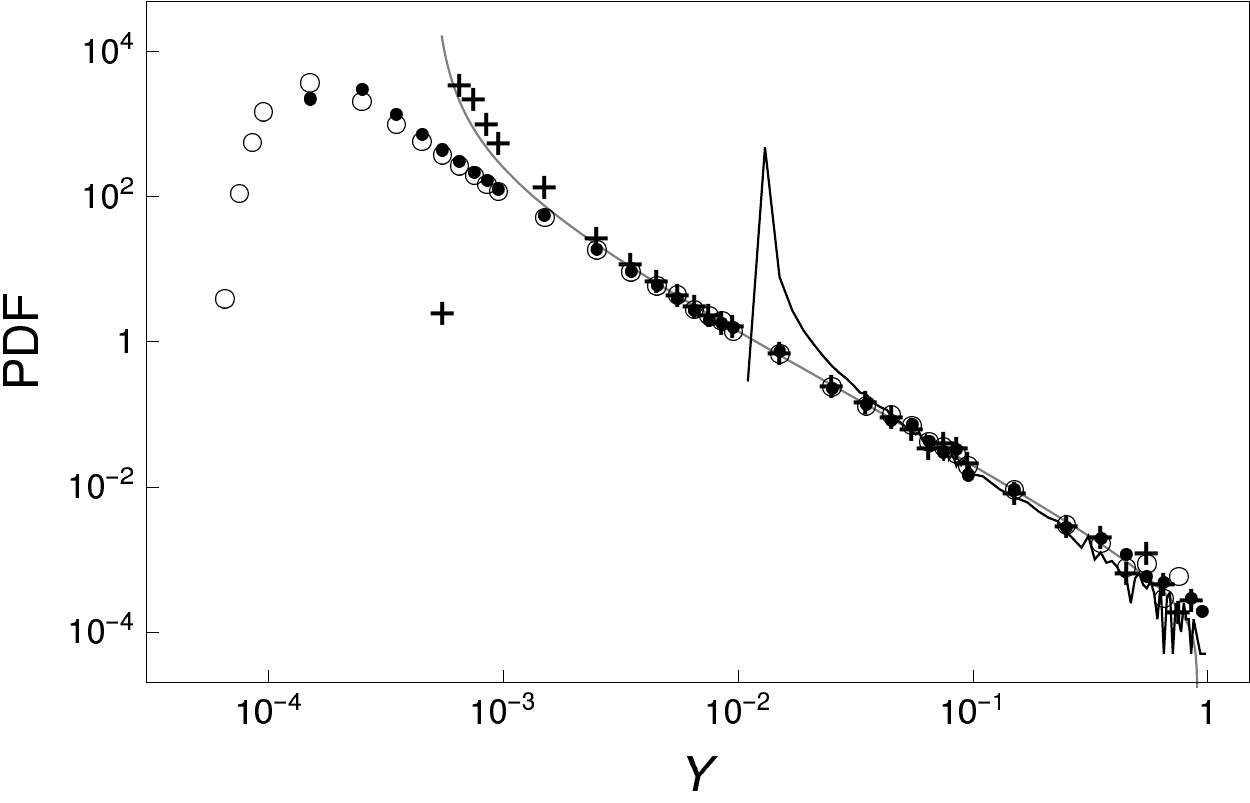}\\
  \small{(c)}&\\
  \includegraphics[height=.302\textwidth,viewport=0 0 360 233]{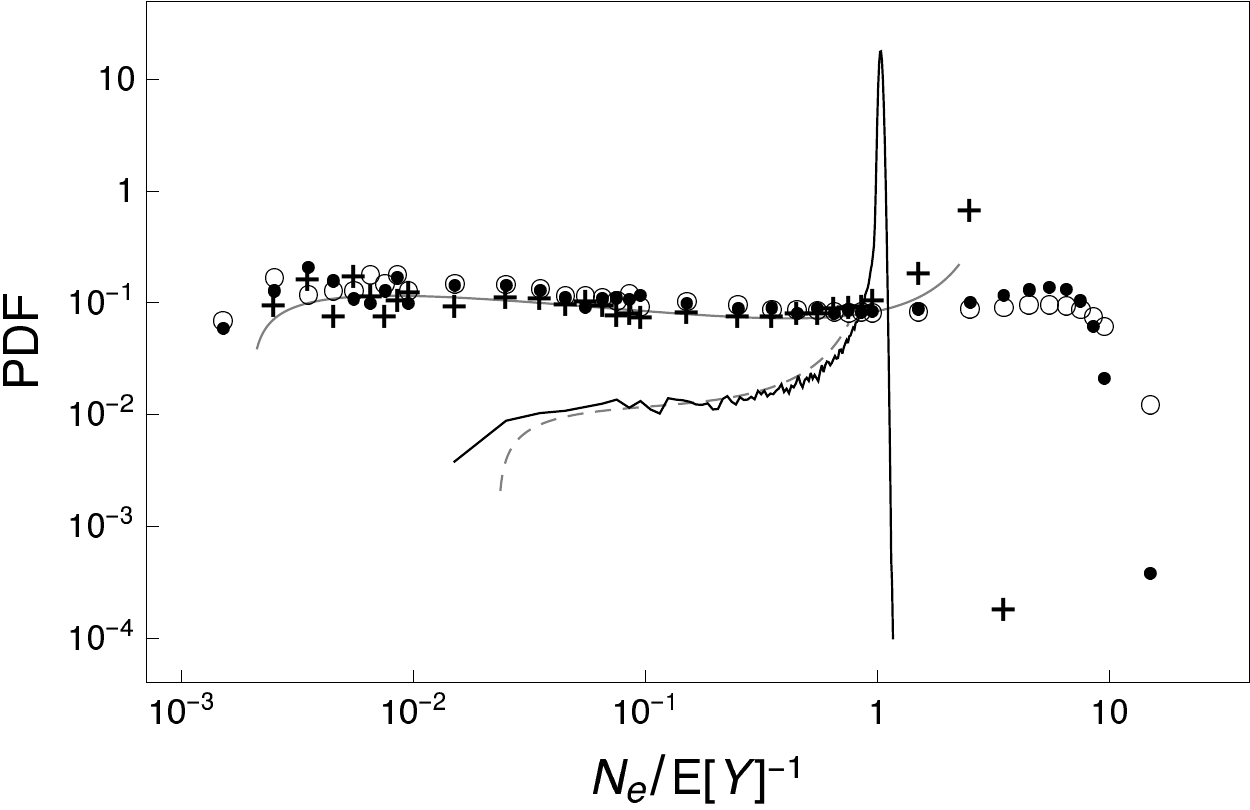}&\\
 \end{tabular}
 \caption{\small
Probability distributions of $W_{1,N}$, $Y$ and $N_{\mathrm{e}}$ in the case $\alpha=1.5$.
The $\mathrm{\Pi}_{W_{1,N}}(w)$ of the largest weight (panel~a),
the $\mathrm{\Pi}_{Y}(y)$ of the average weight (panel~b), and
the $\mathrm{\Pi}_{N_{\mathrm{e}}}(y)$ of the effective population size (panel~c)
are shown for four different settings in Table~\ref{table:simulation-settings}.
Open circles,
solid circles and plus signs are, respectively, obtained by setting~(i), (ii) and~(iii);
the noisy curves are obtained by setting~(iv).
The gray solid lines are from Equations~\eqref{eqn:largest-W-large-N}, \eqref{eqn:distr-Y2-cr} and~\eqref{eqn:distr-Ne-cr} by setting~(iii).
The gray dashed lines in panel~c is from Equation~\eqref{eqn:distr-Ne-yas} by setting~(iv).
$N_{\mathrm{e}}$'s are scaled by $\mathrm{E}[Y]^{-1}$.
 }\label{fig:random-sampling}
\end{figure}

% Table.1
{\small
\renewcommand{\arraystretch}{1.5}
\begin{table}[htb!]
 \caption{Simulation parameter settings}
 \label{table:simulation-settings}
 \begin{tabular}{|c|c|c|c|}
  \hline
  (i) & (ii) & (iii) & (iv) \\
  \hline\hline
  Low specific production & \multicolumn{3}{c|}{High specific hatchery-production}\\
  $p^2/N_1<q^2\hat{w}_N^2$ & \multicolumn{3}{c|}{$p^2/N_1>q^2\hat{w}_N^2$}\\
  \cline{2-4}
  & $p^2/N_1\simeq q^2\hat{w}_N^2$ & $p^2/N_1\simeq c_2$ & $p/N_1>q\hat{w}_N$\\
  \hline
  $N_1=500$ & $N_1=50$ & $N_1=5$ & $N_1=5$\\
  $p=0.05$ & $p=0.05$ & $p=0.05$ & $p=0.25$\\
  \hline
  \multicolumn{4}{l}{
  $q=1-p,{\ } \hat{w}_N\sim {N^{1/\alpha-1}},{\ } N=10^5,{\ } \alpha=1.5$
  }
 \end{tabular}
\end{table}
\renewcommand{\arraystretch}{1.0}
}
The exponent of the $\mathrm{Pareto}(\alpha)$ distribution is set to $\alpha=1.5$.
I generate $\mathrm{\Pi}_{W_{1,N}}(w)$, $\mathrm{\Pi}_{Y}(y)$ and $\mathrm{\Pi}_{N_{\mathrm{e}}}(y)$
by four different settings in Table~\ref{table:simulation-settings}.
The simulation setting~(i) corresponds to the low-productivity case.
The simulation settings~(ii)--(iv) corresponds to the high-productivity case.
The parameters' values in the simulation setting~(iii) are comparable to those of the \textit{Scomberomorus niphonius} population in the Seto Inland Sea, Japan, under hatchery stocking
\citep{Nakajima-Kitada-cjfas2014,Ishida-Katamachi2017}.
The value of $\alpha$ was estimated from analysis of \textit{S. niphonius} mitochondrial DNA sequence variation
using Beta-coalescents (refer to Appendix \ref{app}).
The histograms of $W_{1,N}$'s, $Y$'s and $Y^{-1}$'s,
obtained from $10^5$ independent realizations of $X_j^{(i)}$'s (or $10^6$ realizations for setting~(iv)),
are shown in Figure~\ref{fig:random-sampling}.

% Fig.1
In setting~(i) with $p^2/N_1<q^2\hat{w}_N^2$, the numerical results agree with the asymptotic expressions (Equations~\ref{eqn:largest-W-large-N}, \ref{eqn:distr-Y2}, and~\ref{eqn:distr-Ne}).
The mean value of $Y$ ($1.32\times 10^{-3}$ in the simulation vs. $\mathrm{E}[Y]=1.30\times 10^{-3}$ in Equation~\eqref{eqn:mean-Y2})
is very different from the most probable $Y$ (of order $N^{2(1/\alpha-1)}\sim 10^{-4}$).
The most probable $N_{\mathrm{e}}$ (or the typical reciprocal of $Y$) is at close to the lower end of the $N_{\mathrm{e}}$-distribution, and very different from the reciprocal of the typical $Y$.

In setting~(ii) with $p^2/N_1\simeq q^2\hat{w}_N^2$,
the crossover occurs from violation to restoration of the reciprocal symmetry.
More $Y$'s are concentrated around $p^2/N_1$, so that
the $\mathrm{\Pi}_{N_{\mathrm{e}}}(y)$ has a mode near the upper extreme, as well as a mode at the lower end of the distribution.
The typical $N_{\mathrm{e}}$ has a jump of order $N^{2(1-1/\alpha)}\sim 10^4$.

In setting~(iii) with $p^2/N_1\simeq c_2$,
the numerical results agree with the asymptotic expressions (Equations~\ref{eqn:largest-W-large-N}, \ref{eqn:distr-Y2-cr}, and~\ref{eqn:distr-Ne-cr}).
The mean value of $Y$ ($1.88\times 10^{-3}$ in the simulation vs. $\mathrm{E}[Y]=1.79\times 10^{-3}$ in Equation~\eqref{eqn:nest-site-coal-probab})
is rather close to its most probable value $p^2/N_1=0.5\times 10^{-3}$.

In setting~(iv) with $p/N_1>q\hat{w}_N$, a degeneracy in the probability distribution of $W_{1,N}$ is found around $p/N_1$.
The $Y$ (resp. the $N_{\mathrm{e}}$) shows a peak around $p^2/N_1$ (resp. around $N_1/p^2$).
The numerical results agree with the asymptotic expressions (Equations~~\ref{eqn:largest-W-yas}, \ref{eqn:distr-Y2-yas}, and~\ref{eqn:distr-Ne-yas}).
The mean value of $Y$ ($1.35\times 10^{-2}$ in the simulation)
is very close to its most probable value $p^2/N_1=1.25\times 10^{-2}$.

%{Remark 4}
\begin{remark}
When the average weight of families from the hatchery is increased to a critical value
$p^2/N_1\simeq q^2\hat{w}_N^2$,
a transition takes place:
the $N_{\mathrm{e}}$-distribution changes from unimodal to bimodal,
so that the typical or most probable $N_{\mathrm{e}}$ jumps between the lower and upper extremes of the distribution.
When $p^2/N_1\lesssim q^2\hat{w}_N^2$, the reciprocal symmetry breaking occurs.
This symmetry breaking disappears at the increased specific hatchery-production.
Therefore, the typical $N_{\mathrm{e}}$ divided by the harmonic mean is a convenient order parameter which indicates breaking or restoring the reciprocal symmetry.
\end{remark}

%{Remark 5}
\begin{remark}
In the case $p/N_1\lesssim N^{1/\alpha-1}$,
 fluctuations themselves dictate the main feature of $Y$, while the mean $\mathrm{E}[Y]$ becomes irrelevant for a particular realization or observation (i.e. non-self-averaging effects).
Each realization of the $Y$ (and $N_{\mathrm{e}}$) may be very different from its other realizations.
\end{remark}

\section{Conclusion}
In this paper I have provided the first systematic analysis of the fluctuations of the $N_{\mathrm{e}}$ in the Ryman-Laikre model with hatchery stocking of marine fisheries in mind.
The $N_{\mathrm{e}}$-distribution of the Pareto population ($1<\alpha<2$) with hatchery inputs from a Dirac delta (or gamma) distribution
remains broad even at high specific hatchery-productions and in the large-population limit.
When the hatchery productivity is low,
the most probable $N_{\mathrm{e}}$ is independent of the population size $N$,
close to the lower bound of the distribution
and very different from the harmonic mean of $N_{\mathrm{e}}$'s over replicate populations.
Therefore, the typical average weight of families ($Y\sim N^{2(1/\alpha-1)}$) and its reciprocal (typical $N_{\mathrm{e}}$) do not vary with population size in opposite ways.
When the average weight of families from the hatchery, $p^2/N_1$, is increased to a critical value $q^2\hat{w}_N^2{\;}(\sim N^{2(1/\alpha-1)})$,
the reciprocal symmetry breaking disappears.
The system undergoes a discontinuous transition with the typical $N_{\mathrm{e}}$ jumping from the lower to the upper extreme of the distribution.
At a very high specific hatchery-production
$u\gtrsim\mathrm{E}[X_{1,N_2}^{(2)}]$,
that is, if the weight of a family from the hatchery is greater than the typical weight of the largest family in the wild ($p/N_1\gtrsim q\hat{w}_N$),
the population is swamped with $N_1$ families from the hatchery.

Under the assumption of a Pareto offspring-number distribution,
the potential practical consequences are dramatic.
There is an inevitable deviation between the Ryman-Laikre formula prediction and the observed effective size of the admixed population.
There will be observed a lack of the Ryman-Laikre effect.

\appendix
\setcounter{figure}{0}
\renewcommand{\thefigure}{A\arabic{figure}}

\section{\texorpdfstring{\underline{S. niphonius}}{TEXT}
in the Seto Inland Sea, Japan}\label{app}
I used the published data from the Seto Inland Sea (SIS) stock of Japanese Spanish mackerel \textit{Scomberomorus niphonius}
\citep{Nakajima-Kitada-cjfas2014}.
I studied variation in partial mtDNA control region sequences of 330 wild mature fish collected in 2010;
a total of 63 sequence types
(53--56, 59, 61, 63--65, 68--72, 77--79, 81--83, 85, 87, 89, 90, 92--01, 04, 05, 07--09, 14, 15, 19, 21--26, 28, 30--32, 34--40, 43--46)
were retrieved from GenBank
(the last two digits of haplotypes with the GenBank accession numbers AB844453--AB844546).

\subsection{Remedying infinitely-many-sites model violations}
\begin{figure}[htb]
   \centering
    \begin{tabular}{ll}
     \small{(a)} & \small{(b)}\\
     \includegraphics[height=.46\textwidth,viewport=0 0 360 358]{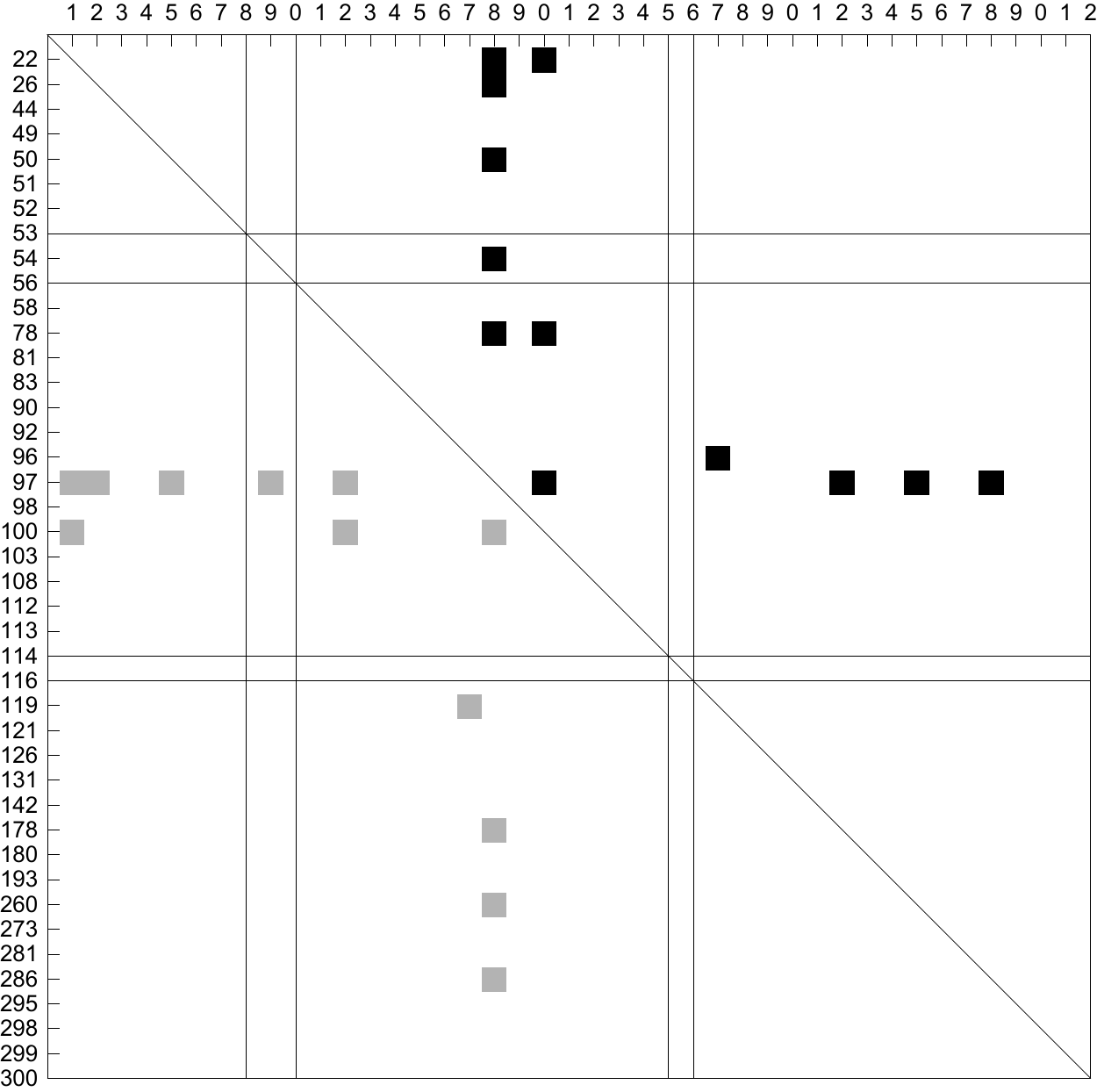}&
     \includegraphics[height=.46\textwidth,viewport=0 0 360 358]{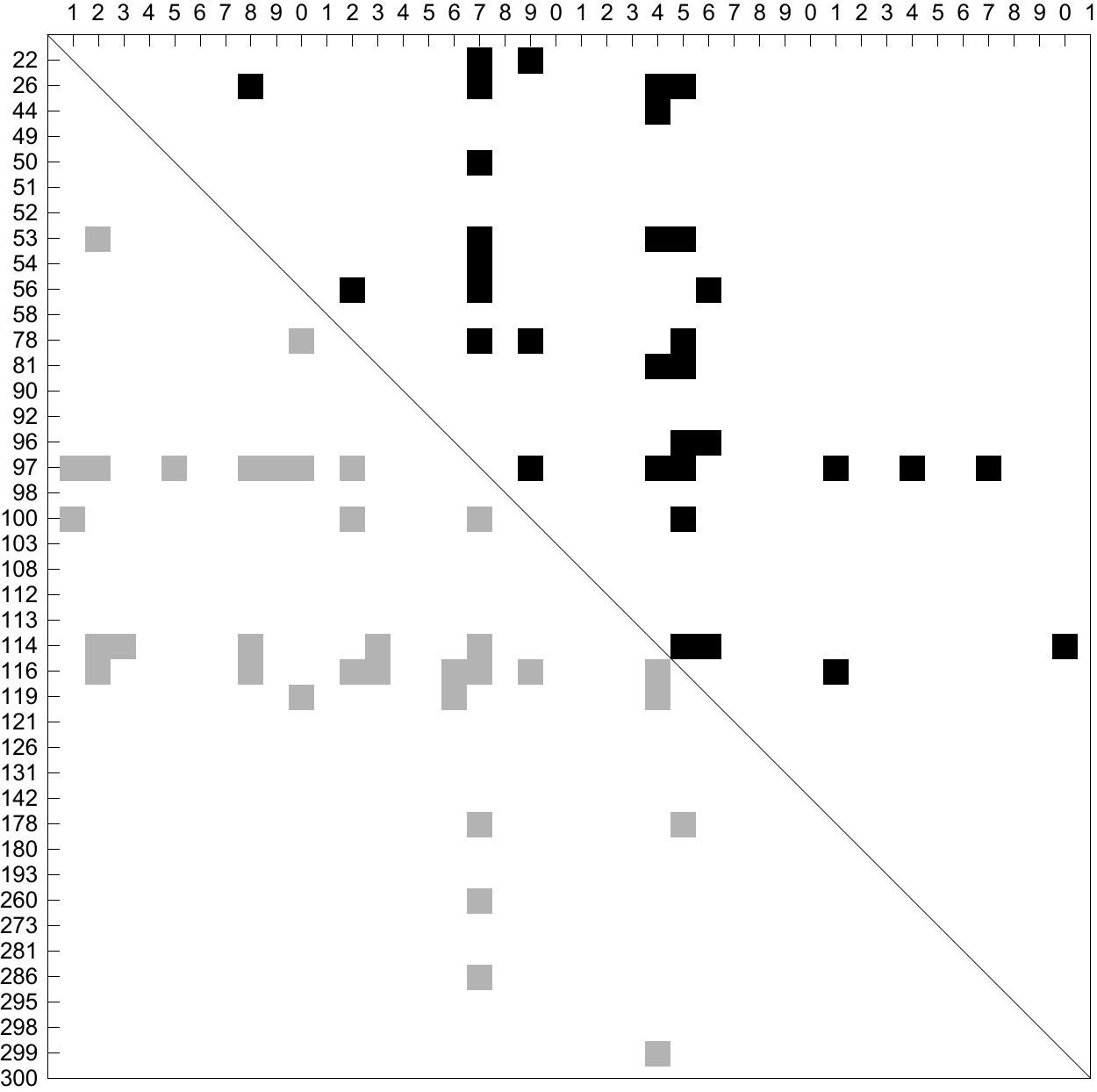}
    \end{tabular}
    \caption{\small
   Compatibility matrix for the \textit{S. niphonius} mtDNA CR sequences (partial 305 base pairs) from the SIS sample in 2010.
    Numbers in left column designate variable sites in the sequences.
    (a)~There are 38 biallelic and 4 triallelic sites (sample size 330); the horizontal (and vertical) lines indicate the triallelic sites.
     The diagonal line highlights the symmetry in the matrix.
    (b)~After removal of six sequence types that ten individuals possess,
    no triallelic sites are retained in sequence data.
    There are 41 segregating sites (sample size 320).
    }\label{compatibility_matrix}
  \end{figure}

I assumed the simplest possible substitution process, the infinitely-many-sites model (ISM) of neutral mutation \citep{Watterson75}.
Since mutations can occur only once at a given site, there are an ancestral type and a mutant type at each segregating site.
So, 0 and 1 just represent two types of nucleotide base.
A pair of polymorphic sites are compatible with each other, if there are fewer of the four possible combinations of state (00, 01, 10 and 11) occurred in two columns of the alignment.
If all four combinations of 0 and 1 are present in two columns, this pattern shows the violation of the ISM, and the sequences cannot be represented by a unique gene tree \citep{Griffiths2001}.
The program \texttt{genetree} \citep{Griffiths2001} was used to report on the inconsistencies in the data with regards to the ISM.
The compatibility matrix (Figure~\ref{compatibility_matrix}a) computed for the data from 330 individuals examines the overall support or conflict among variable sites in the mtDNA CR sequences.
The upper triangle checks for broad incompatibility (indicated by black), the lower triangle for narrow incompatibility (gray, where the commoner of the two alleles was taken to be ancestral), and compatible sites are left white.
Narrow incompatibility (incompatible in a rooted sense) means that this site is only a problem with the current arrangement of 0's and 1's for this particular rooted tree with the assumption that 0 is the ancestral type.
Changing 0's to 1's in either column (this produces a different rooted tree) will make the data ISM compatible and able to be turned into a unique tree.
Broad incompatibility (incompatible in an unrooted sense) means that even toggling the 0 and 1's at each segregating site does not make this data set consistent with the ISM and a gene tree cannot be produced.

% Fig.A1

I removed incompatible sites in an unrooted sense from the mtDNA CR sequences, without specifying which of the two alleles is the oldest,
and found the largest set of sites that is consistent with the ISM.
There were 31 segregating sites defining 30 haplotypes from 330 individuals
(Figures~\ref{compatibility_matrix}a and~\ref{mgtree-beta}a).
This treatment is not likely to bias the analyses,
because of the thinning property of Poisson random variables \citep{Kingman93} that removing points randomly from the original Poisson point process results in another Poisson point process with the remaining points.
It is therefore assumed that,
conditional on the ancestral tree of a subset of sites compatible with the ISM, mutations occur at the points of Poisson processes of rate $\theta/2$, independently on each branch of the tree,
where time (branch length) is measured in coalescent time units.

After removal of six sequence types with GenBank accession numbers {55, 97--99, 09, 39} that ten individuals possess
(which removes triallelic sites in sequence data),
computing the compatibility matrix (Figure~\ref{compatibility_matrix}b)
I solved the violations of the ISM by excluding topologically incompatible sequences
{53, 54, 77, 78, 82, 85, 90, 92, 00, 05, 07, 08, 21, 23, 24, 28, 31, 32, 34, 35, 38, 40, 43--46}
(the last two digits of haplotypes with the GenBank accession numbers),
where I selected the largest subset of sequences to which the ISM applied,
resulting in a total of 34 segregating sites defining 31 haplotypes from 285 individuals (Figure~\ref{mgtree-beta}c).

\subsection{Rooted gene tree}
The rooted gene tree is a condensed description of the coalescent tree and shows the ancestral relationships between genes.
Taking into account the topology of the tree, the root was chosen by likelihood under the Kingman coalescent model.
I ran $10^5$ repetitions of the simulation algorithm \citep{Griffiths-Tavare94} to find the likelihood of each of the possible rooted trees as a function of the population-scaled mutation rate $\theta$,
summed these to find the probability of the unrooted tree,
and from this I obtained a maximum likelihood estimate (MLE) of $\theta$.
Likelihood calculations were carried out with {\texttt{genetree}} program.

% Fig.A2
  \begin{figure}[htb]
\centering{
   \begin{tabular}{ll}
    \small{(a)}&\hspace{-17mm}\small{(b)}\\
     \hspace*{-11mm}
    {\includegraphics[width=.665\textwidth,viewport=0 0 503 204]{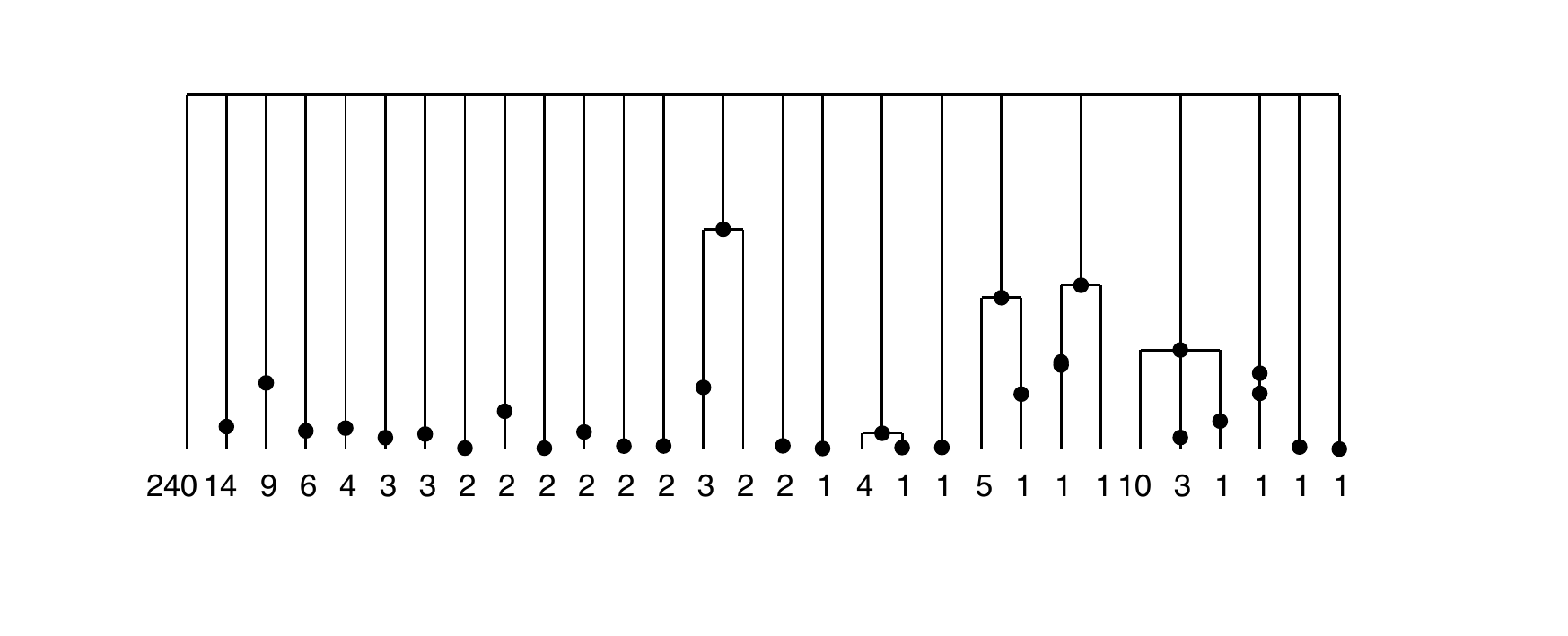}}&
         \hspace{-14mm}
    {\includegraphics[width=.405\textwidth,viewport=0 0 360 359]{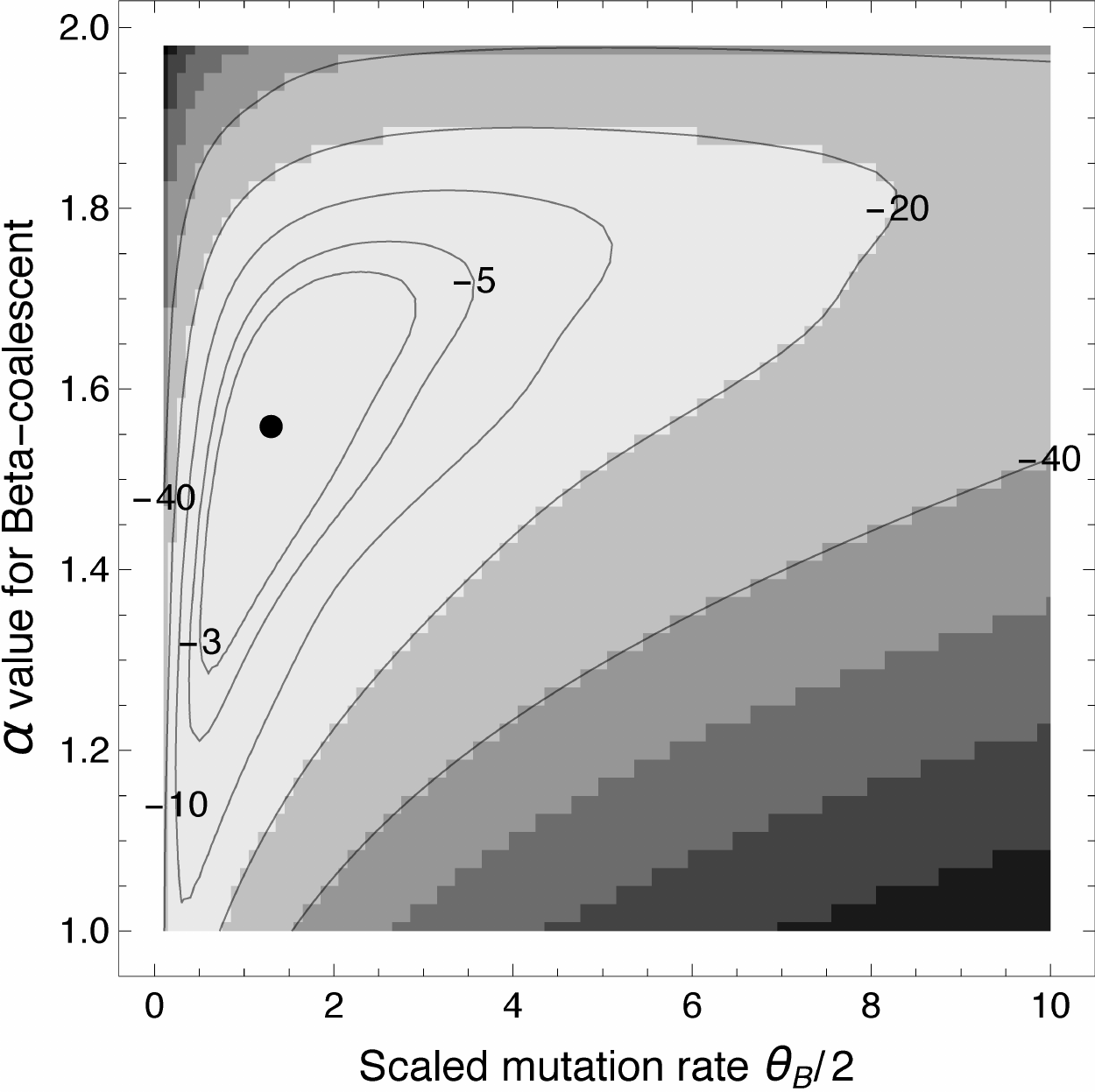}}\\
    \small{(c)}&\hspace{-17mm}\small{(d)}\\
    \hspace*{-11mm}
    {\includegraphics[width=.665\textwidth,viewport=0 0 502 204]{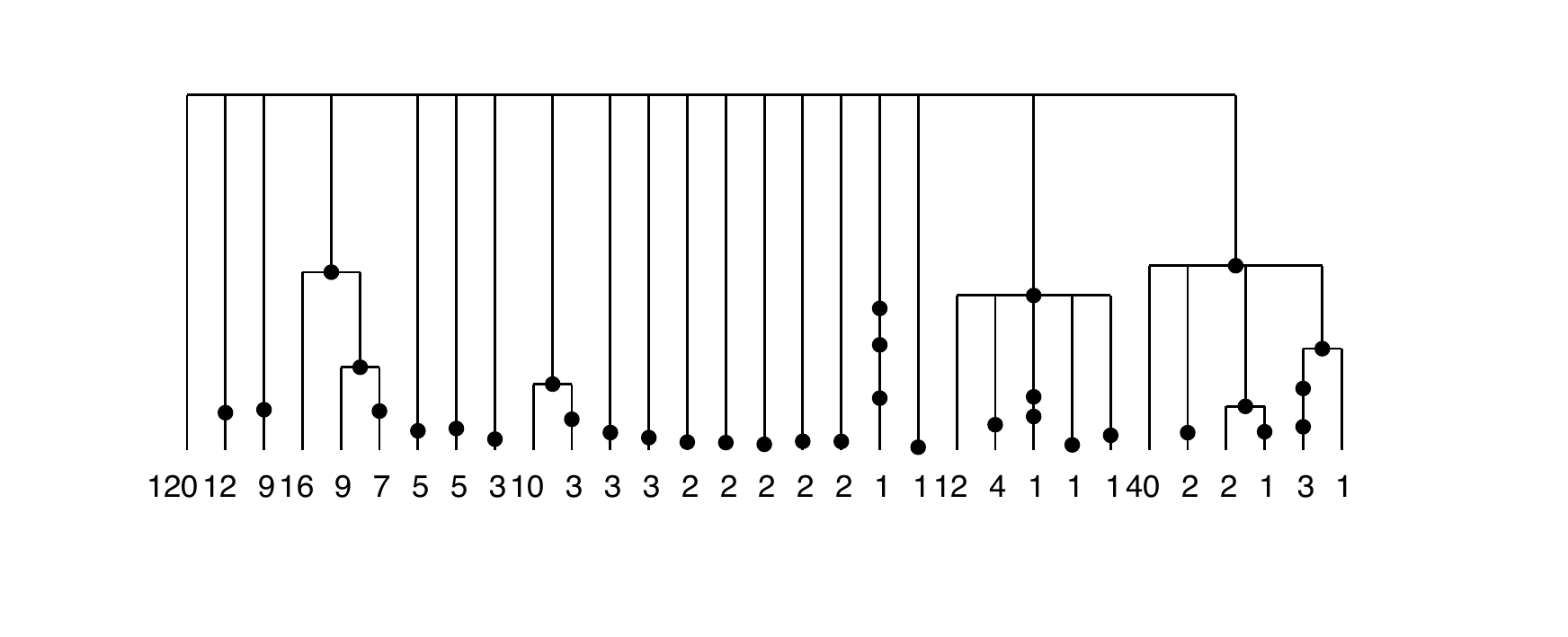}}&
        \hspace{-14mm}
    {\includegraphics[width=.405\textwidth,viewport=0 0 360 359]{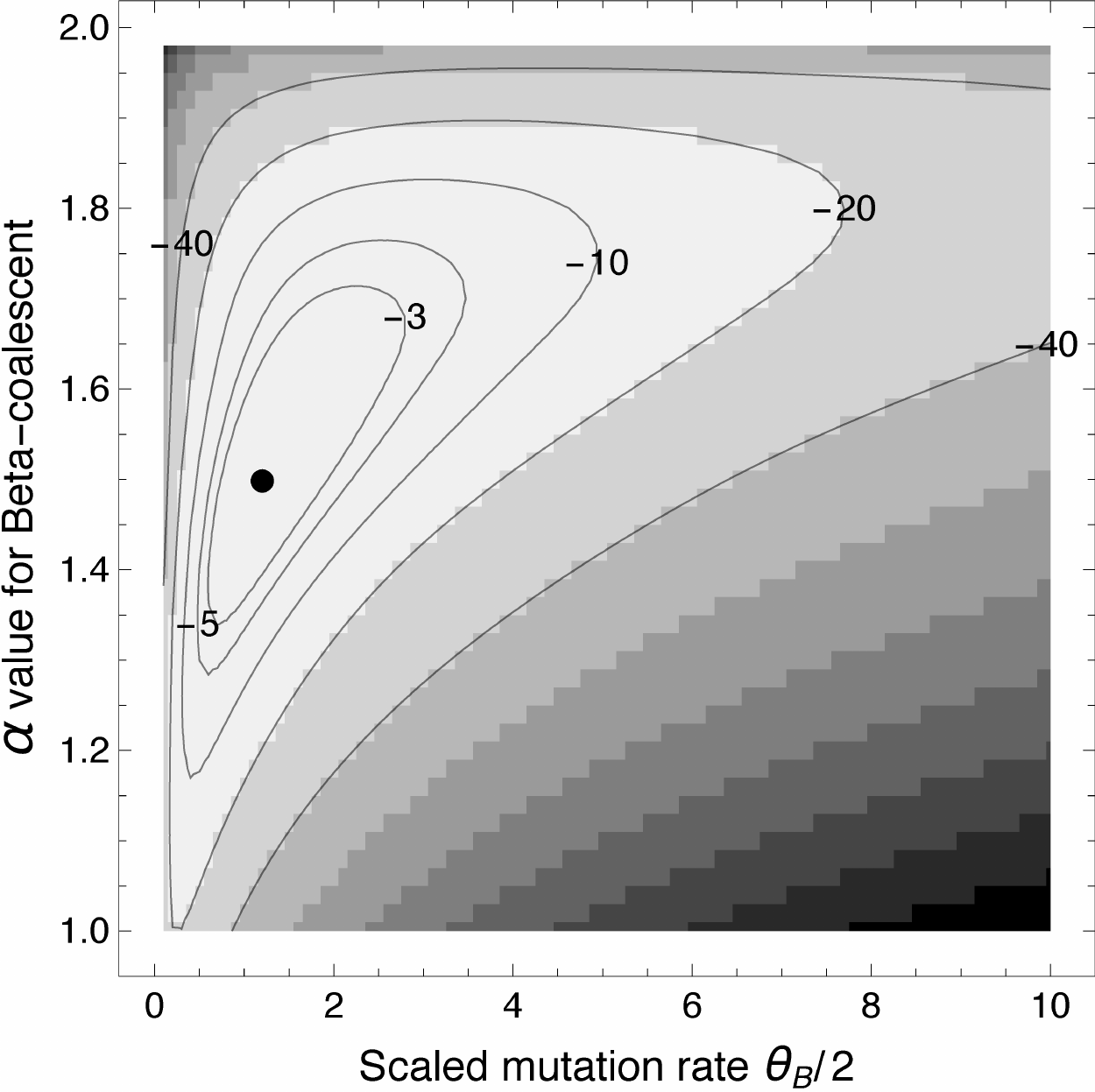}}
   \end{tabular}
   \caption{\small
Coalescent analysis of the \textit{S. niphonius} mtDNA CR sequences from the SIS sample in 2010.
   (a)~Maximum likelihood rooted gene tree for 330 sequences excluding topologically incompatible sites.
   There were 31 segregating sites defining 30 haplotypes.
   The dots indicate mutations.
   The numbers beneath the tree tips give the multiplicity of each unique haplotype.
   (b)~Log-likelihood surface (scaled to maximum log likelihood $-36.3$) for the ML rooted gene tree (panel a)
   analyzed under the {$\mathrm{Beta}(2-\alpha,\alpha)$} coalescent.
   The arg-maximum of the likelihood surface is indicated by a dot $(\theta_{\mathrm{B}}/2=1.3,{\,}\alpha=1.56)$.
   (c)~ML rooted gene tree obtained with the largest ISM-compatible subset ($n=285$) of sequences.
   There were 34 segregating sites defining 31 haplotypes.
   (d)~Log-likelihood surface (scaled to maximum log likelihood $-77.8$) for the ML rooted gene tree (panel c).
   The arg-maximum of the likelihood surface is
   $(\theta_{\mathrm{B}}/2=1.2,{\,}\alpha=1.50)$.
   The 95\% joint confidence contour (likelihood based) is defined by taking the values of {$(\theta_{\mathrm{B}}/2,\alpha)$} for which the natural logarithm of the likelihood is {$\chi_{0.95}^2(2)/2=2.995$} units smaller than the log of maximum likelihood, where {$\chi_{0.95}^2(2)$} is the 0.95 quantile of a {$\chi^2$}-distribution with two degrees of freedom.
    }\label{mgtree-beta}
   }
  \end{figure}

After removing incompatible sites from the \textit{S. niphonius} mtDNA CR sequences, there are 31 segregating sites and thus, there are 32 possible rooted trees.
Using the MLE of $\theta=5.4$ from the unrooted tree
the probabilities (relative likelihoods) of the 32 rooted trees were estimated, which varied between $4.48\times 10^{-7}$ and $0.990$.
Figure~\ref{mgtree-beta}a shows the rooted gene tree with the highest probability $0.990$.

In the largest ISM-compatible subset ($n=285$) of sequences from the 2010 sample, there are 34 segregating sites and thus, there are 35 possible rooted trees.
Using the MLE of $\theta=10.7$ from the unrooted tree
the probabilities (relative likelihoods) of the 35 rooted trees were estimated, which varied between $2.42\times 10^{-12}$ and $0.99999$.
Figure~\ref{mgtree-beta}c shows the rooted gene tree with the highest probability $0.99999$.
The gene trees, representing the mutation paths to the root, were produced using {\texttt{genetree}} program.
Note that the most frequent allele at each site coincides with the oldest.

\subsection{Beta coalescents}
Consider the $\mathrm{Beta}(2-\alpha,\alpha)$ coalescent as the underlying model, which yields a coalescent history that is consistent with the \textit{S. niphonius} mtDNA gene trees (Figures~\ref{mgtree-beta}a and~c),
where mutations appear along the branches at rate $\theta_{\mathrm{B}}/2$ according to the ISM.
I computed the likelihoods of the rooted gene trees under the parameters $(\theta_{\mathrm{B}}/2,\alpha)$ employing an importance sampling scheme using program \texttt{MetaGeneTree}
\citep{BBS11},
which is an extension of the method \citet{Griffiths-Tavare94} developed for the Kingman coalescent.
Likelihood values were estimated independently for each discrete gridpoint using $10^6$ independent runs of the Markov chain with spacing $(\Delta\theta_{\mathrm{B}}/2,\Delta\alpha)=(0.1,0.02)$ between gridpoints.
For 330 sequences excluding topologically incompatible sites,
the MLEs were $\theta_{\mathrm{B}}/2=1.3$ and $\alpha=1.56$.
For the largest ISM-compatible subset ($n=285$) of sequences,
the MLEs were $\theta_{\mathrm{B}}/2=1.2$ and $\alpha=1.50$.
It turned out that
the difference between the two estimates is statistically not significant
(Figures~\ref{mgtree-beta}b and~d), as was expected.
The value of the coalescent parameter $\alpha<2$ was significant.

%REFERENCES
\bibliographystyle{apalike2}
\bibliography{bib-niwa-genetics}

\end{document}